\documentclass[11pt,a4paper]{article}
\usepackage{jheppub}

\usepackage{bbm}
\usepackage{amsfonts}
\usepackage{booktabs}
\usepackage{mathrsfs}
\usepackage{epsfig}
\usepackage{graphicx}
\usepackage{dcolumn}
\usepackage{bm}
\usepackage{amsmath}
\usepackage{slashed}

\let\jnfont=\rm
\def\NPB#1,{{\jnfont Nucl.\ Phys.\ B }{\bf #1},}
\def\PLB#1,{{\jnfont Phys.\ Lett.\ B }{\bf #1},}
\def\EPJC#1,{{\jnfont Eur.\ Phys.\ Jour.\ C }{\bf #1},}
\def\PRD#1,{{\jnfont Phys.\ Rev.\ D }{\bf #1},}
\def\PRL#1,{{\jnfont Phys.\ Rev.\ Lett.\ }{\bf #1},}
\def\MPLA#1,{{\jnfont Mod.\ Phys.\ Lett.\ A }{\bf #1},}
\def\JPG#1,{{\jnfont J.\ Phys.\ G}{\bf #1},}
\def\CTP#1,{{\jnfont Commun.\ Theor.\ Phys.\ }{\bf #1},}
\def\ZPC#1,{{\jnfont Z.\ Phys.\ C }{\bf #1},}
\def\JHEP#1,{{\jnfont JHEP \ }{\bf #1},}
\def\Rv{\not{\hbox{\kern-1pt $R$}}}
\def\p{\not{\hbox{\kern-3pt $p$}}}

\title{SUSY effects in Higgs productions at high energy $e^+e^-$  colliders}

\author{ Junjie Cao$^1$, Chengcheng Han$^2$, Jie Ren$^1$, Lei Wu$^3$, Jin Min Yang$^4$, Yang Zhang$^4$}

\affiliation{
$^1$ Physics Department, Henan Normal University,
     Xinxiang 453007, China \\
            $^2$Asia Pacific Center for Theoretical Physics,
             San 31, Hyoja-dong, Nam-gu,
             Pohang 790-784, Republic of Korea \\
 $^3$ ARC Centre of Excellence for Particle Physics at the Terascale, School of Physics,
      University of Sydney, NSW 2006, Australia\\
 $^4$  State Key Laboratory of Theoretical Physics,
       Institute of Theoretical Physics,
       Academia Sinica, Beijing 100190, China
    }

\abstract{
Considering the constraints from collider experiments and dark matter detections,
we investigate the SUSY effects in the Higgs productions $e^+e^- \to Zh$ at an
$e^+e^-$ collider with a center-of-mass energy above 240 GeV
and $\gamma\gamma \to h \to b\bar{b}$ at a photon collider with
a center-of-mass energy above 125 GeV.
In the parameter space allowed by current experiments, we find that the SUSY corrections
to $e^+e^- \to Zh$ can reach a few percent and the production rate of
$\gamma\gamma \to h \to b\bar{b}$ can be enhanced by a factor of 1.2 over the SM prediction.
We also calculate the exotic Higgs productions $e^+e^-\to Zh_1$ and
$e^+e^-\rightarrow A_1h$ in the next-to-minimal supersymmetric model (NMSSM)
($h$ is the SM-like Higgs, $h_1$ and  $A_1$ are
respectively the CP-even and CP-odd singlet-dominant Higgs bosons which can be much lighter
than $h$). We find that at a 250 GeV $e^+e^-$ collider
the production rates of $e^+e^-\rightarrow Zh_1$ and $e^+e^-\to A_1h$
can reach 60 fb and 0.1 fb, respectively.
}

\begin{document}
\maketitle \indent

\section{Introduction}
The LHC has discovered a scalar with mass around 125 GeV which resembles the Standard Model
(SM) Higgs boson \cite{ATLAS-CMS}. Since the minimal supersymmetric model (MSSM)
predicts a light Higgs boson below 130 GeV \cite{effective-potential}, the discovery of
such a 125 GeV Higgs boson may be the first hint of low energy supersymmetry (SUSY).
However, the LHC measurements of the properties of this new boson are so far consistent
with the SM predictions, which squeezes the SUSY effects in the Higgs couplings
to a decoupling region \cite{Cao,belanger,gardino}.
Besides, after the LHC Run-1, the null results of direct searches for SUSY particles
(sparticles) have excluded the first two generation squarks and gluino with mass
below about 1 TeV \cite{AbdusSalam}. The third generation squarks and non-colored
sparticles as light as hundreds of GeV are still allowed but have also been
constrained by the LHC searches \cite{stop-gauginos}.
All these indicate that the SUSY scale may be
much higher than the electroweak scale. So it will be a challenge for the LHC to
directly observe any SUSY particles except for the alone light Higgs boson.
In such a situation, an alternative way for probing SUSY is to search for the
indirect SUSY loop effects from some precision measurements of the Higgs boson.
Since the precision measurements of the Higgs boson
are rather challenging at hadron colliders like
the LHC, some high energy $e^+ e^-$ colliders with center-of-mass energy above 240 GeV
are being proposed.

At an $e^+ e^-$ collider, the Higgs-strahlung process $e^+e^- \to Zh$ is the dominant
production channel for the Higgs boson, for which the $Zh$ events can be inclusively
detected by tagging a leptonic $Z$ decay without assuming the Higgs decay mode.
For a center-of-mass energy of $240-250$ GeV and an integrated luminosity of 500 fb$^{-1}$,
 an $e^+ e^-$ collider can produce about $O(10^5)$ Higgs bosons per year and
allow for measuring the Higgs couplings at percent level \cite{peskin,Higgsfactory},
which may be able to unravel the SUSY effects in this production.
For this process, the leading order rate, the one-loop electroweak corrections
and the SUSY corrections were calculated in \cite{zh-tree},
\cite{zh-loop1,zh-loop2,zh-loop3,zh-real} and \cite{zh-susy}, respectively.

As a feasible option, the $\gamma\gamma$ collision can be achieved through the backward
Compton scattering of laser light against high-energy electrons at a linear $e^+e^-$ collider.
At such a  $\gamma\gamma$ collision the Higgs boson can be singly produced via the loop
process $\gamma\gamma \to h$. This process is demonstrated to be sensitive to the
new charged SUSY particles. So the photon collider will be an ideal place to investigate
the anomalous $h\gamma\gamma$ coupling. At the $\gamma\gamma$ collider, the Higgs partial
width $\Gamma_{\gamma\gamma}$ can be measured with an accuracy of about 2\%.
Besides, the $CP$ property of the Higgs boson can be measured using the photon polarizations.
The single production of SUSY Higgs bosons through $\gamma\gamma$ fusion has been calculated
in \cite{rrh-mssm-cx,rrh-mssm-cp}.

Note that at a high energy  $e^+ e^-$ collider the  productions of some exotic Higgs
bosons will be possible. If the center-of-mass energy is designed at 240-250 GeV,
the production  $e^+e^- \to h A$ in the MSSM, which is complementary to the production
 $e^+e^- \to Zh$ and was searched at LEP2, will not be open because the CP-odd Higgs
$A$ is now much heavier than the SM-like Higgs $h$. However, in the NMSSM the lightest
CP-even Higgs $h_1$ and CP-odd Higgs $A_1$ can be singlet-dominant and much lighter
than the SM-like Higgs $h$ \cite{cao-light-scalar}. So in the NMSSM the exotic Higgs productions
$e^+e^-\to Zh_1$ and $e^+e^-\to h A_1$ may occur at a 240-250 GeV $e^+ e^-$ collider.
These exotic Higgs productions could be a good probe for non-minimal SUSY like the
NMSSM.

In this work we examine systematically all the above processes in SUSY.
We will not only calculate the NMSSM processes $e^+e^-\to Zh_1$ and $e^+e^-\to h A_1$
which have not been intensively studied in the literature, but also re-examine
the SUSY effects in $e^+e^- \to Zh$ and $\gamma\gamma \to h$ by considering
current experimental constraints, such as the LHC Higgs data and the dark matter
detection limits.

This work is organized as follows. In Sec. II we describe the parameter scan and the
calculation details for the processes $e^+ e^- \to Zh$, $\gamma\gamma \to h \to b\bar{b}$,
$e^+e^-\to Zh_1$ and  $e^+e^-\to A_1h$. In Sec.III, we show the numerical results.
Finally, we draw the conclusions in Sec. IV.

\section{Calculations}
\subsection{A description of calculations }
There are about 120 free parameters in a general $R$-parity conserving weak-scale MSSM.
However, most of these parameters are related to the flavor changing neutral currents (FCNC)
and/or the $CP$-violating phases, which are highly constrained by the experimental
measurements. So in our work we only discuss the pMSSM and CMSSM, where the free parameters
are reduced and the models are more predictive.

The pMSSM is considered as the most general version of the $R$-parity conserving MSSM with
the following considerations
\begin{itemize}
\item[(i)] $CP$ conservation;
\item[(ii)] The principle of minimal flavor violation (MFV) at the weak scale;
\item[(iii)] Degenerate masses of the first and second generation sfermions;
\item[(iv)] Negligible Yukawa couplings and trilinear terms for the first two generations,
but keeping the 3rd generation parameters $A_t, A_b, A_{\tau}$;
\item[(v)] The lightest neutralino as the LSP.
\end{itemize}
Finally, only 19 parameters can be independently changed in the pMSSM, which are
\begin{itemize}
\item[(a)] $\tan\beta$, which is the ratio of the vevs of the two  Higgs doublet fields;
\item[(b)] the higgsino mass parameter $\mu$ and the pseudo-scalar Higgs mass $m_A$;
\item[(c)] the gaugino mass parameters  $M_1, M_2, M_3$ ;
\item[(d)] the first/second generation sfermion mass parameters
           $m_{\tilde{q}}, m_{\tilde{u}_R}, m_{\tilde{d}_R}, m_{\tilde{l}}, m_{\tilde{e}_R}$;
\item[(e)]  the third generation sfermion mass parameters
            $m_{\tilde{Q}}, m_{\tilde{t}_R}, m_{\tilde{b}_R}, m_{\tilde{L}}, m_{\tilde{\tau}_R}$;
\item[(f)]  the third generation trilinear couplings $A_t, A_b, A_{\tau}$.
\end{itemize}
To further simplify the parameter space we assume $M_1:M_2=1:2$,
$A_t=A_b=A_\tau=A_e$ and $m_{t_R}=m_{b_R},m_{\tau_R}=m_{e_R},m_l=m_L$.
We also take a common mass $M_{SUSY}=m_q=m_{u_R}=m_{d_R}=M_3=2$ TeV to avoid the constraints
from the first-two generation squarks and the gluino direct searches at the LHC.
In addition, considering the current bounds on the stop and stau masses in the MSSM,
we conservatively require the lighter stop mass $m_{\tilde{t}_1}>300$ GeV and the lighter
stau mass $m_{\tilde{\tau}}>150$ GeV. We scan the parameters space in the following ranges
\begin{eqnarray}
  && |A_{t}| \le 5 {\rm ~TeV}, \quad 100 {\rm ~GeV} \le (M_{Q}, M_{t_R}) \le 3 {\rm ~TeV},
\nonumber \\
  && 1 \le \tan\beta \le 50, \quad 90 {\rm ~GeV} \le M_A \le 1.5 {\rm ~TeV},  \nonumber \\
  && 100 {\rm ~GeV} \le \mu \le 1{\rm ~TeV},
\quad 100 {\rm ~GeV} \le (M_{L}, M_{l}, M_2) \le 3~{\rm TeV}.
\end{eqnarray}
Different from the general MSSM where all soft breaking parameters are independent \cite{pmssm},
the CMSSM \cite{cmssm} assumes the following universal soft breaking parameters at SUSY
breaking scale (usually chosen as the Grand Unification scale) as the fundamental ones:
\begin{equation}
  \ M_{1/2} \ , \ M_0 \ , \ A_0 \ , \tan\beta \ , \ {\rm sign}(\mu),
\end{equation}
with $M_{1/2}$, $M_0$ and $A_0$ denoting the gaugino mass, scalar mass and trilinear
interaction coefficient, respectively.
When evolving these parameters down to weak scale, we get all the soft breaking
parameters in the low energy MSSM. The ranges of these parameters in our scan are
\begin{eqnarray}
   && 200 {\rm ~GeV} \le m_0 \le 4 {\rm ~TeV},  \nonumber \\
   &&¡¡200 {\rm ~GeV} \le m_{1/2} \le 4 {\rm ~TeV},  \nonumber \\
 ¡¡&& -4 {\rm ~TeV} \le A_0 \le 4 {\rm ~TeV}, \nonumber \\
   && 2 \le \tan\beta \le 65.
\end{eqnarray}
For the NMSSM, we scan the parameters in the following ranges
\begin{eqnarray}
  && |A_{t}| \le 3~ {\rm TeV}, \quad 100~ {\rm GeV} \le (M_{Q}, M_{t_R}) \le 2 ~{\rm TeV},  \nonumber \\
  && 1 \le \tan\beta \le 10, \quad 90 {\rm GeV} \le M_A \le 1 {\rm TeV}, \nonumber \\
  && 100 ~{\rm GeV} \le \mu \le 300~{\rm TeV}, \quad 100~ {\rm GeV} \le (M_{L}, M_{l}, M_2) \le 3~{\rm TeV}.  \nonumber \\
  &&  0.001 \le \lambda \le 0.8 , 0.001 \le \kappa \le 0.8, \quad  -300 ~{\rm GeV} \le  A_{\kappa} \le 300~{\rm GeV}
\end{eqnarray}
and other parameters like first/second generation squark mass
and gluino mass are set to be 2 TeV.

In our scan, we impose the following constraints
\begin{itemize}
  \item[(1)]
  A SM-like Higgs mass in the range of 123-127 GeV.
    We use \textsf{FeynHiggs-2.8.9} \cite{feynhiggs} to calculate the Higgs mass and impose the experimental constraints from LEP, Tevatron and LHC with \textsf{HiggsBounds-3.8.0} \cite{higgsbounds}. We do not perform the Higgs couplings fit to the LHC data because of the current poor precisions.
  \item[(2)]
    Various $B$-physics bounds at 2$\sigma$ level.
    We implement the constraints by using the package of \textsf{SuperIso v3.3} \cite{superiso} , including $B \to X_s\gamma$ and the latest measurements of $B_s\rightarrow \mu^+\mu^-$, $B_d \to X_s\mu^+\mu^-$ and $ B^+ \to \tau^+\nu$.
   \item[(3)]
    The thermal relic density of the lightest neutralino is in the 2$\sigma$ range of the Planck data \cite{planck} and the dark matter $\sigma^{SI}$ upper limit form the LUX data\cite{LUX}.
    The code \textsf{MicrOmega v2.4} \cite{micromega} are used to calculate the relic density.
    \item[(4)]
    The constraints from the electroweak observables such as
 $\rho_l$, $\sin^2 \theta_{eff}^l$, $m_W$ and $R_b$ \cite{rb} at $2\sigma$ level.
  \item[(5)]
    We require the MSSM and NMSSM to explain at $2\sigma$ level the discrepancy of the measured value
of the muon anomalous magnetic moment from its SM prediction,
i.e., $a^{exp}_{\mu}-a^{SM}_{\mu}$ = $(28.7\pm8.0)\times10^{-10}$ \cite{g-2}.
    While for the CMSSM, since there is a tension between $\mu_{g-2}$ with the Higgs
    mass \cite{cao-cmssm},
    we just require the CMSSM prediction not worse than the SM value.
  \item[(6)]
  Since the large mixing terms in the stop/stau sector will affect the vacuum stability,
we require SUSY to comply with the vacuum meta-stability condition by using the formulas
in \cite{mczhang,jap}.
\end{itemize}
We also impose the multi-jets direct search limits \cite{multi-jets} on the
$(m_0,m_{1/2})$ plane based on the search by the ATLAS collaboration for
squarks and gluinos in the final states that contain missing $E_T$, jets and
$0 - 1$ leptons in $20.1-20.7fb^{-1}$ integrated luminosity of data at $\sqrt{s}=8$ TeV
collision energy. While these exclusion limits were obtained in the MSUGRA/CMSSM framework
for fixed value of $tan\beta$ and $A_0=-2m_0$, it was proved \cite{Allanach:2011ut} that
the result is fairly insensitive to $tan\beta$ and $A_0$ and so we can use the limits directly.

Besides, to make the SM-like Higgs not deviate from the Higgs data largely, we also impose
following constraints on the property of the SM-like Higgs
\begin{eqnarray}
&& 0.8 \leq \frac{\sigma(pp\rightarrow h)\times Br(h\rightarrow \gamma\gamma)}{(\sigma\times Br)_{SM}} \leq 1.5 \\
&& 0.8 \leq \frac{\sigma(pp\rightarrow h) \times Br(h\rightarrow WW/ZZ)}{(\sigma\times Br)_{SM}} \leq 1.2
\end{eqnarray}
In the calculations, we generate and simplify the amplitudes by using
the packages \textsf{FeynArts-3.9} \cite{feynarts} and \textsf{FormCalc-8.2}\cite{formcalc}.
All the loop functions are numerically calculated with the package \textsf{LoopTools-2.8} \cite{looptools}.

\subsection{Calculations for $e^+e^- \to Zh$}
\begin{figure}[t]
 \includegraphics[width=15cm]{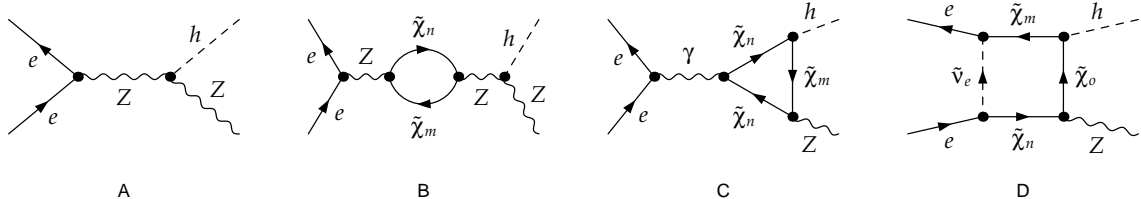}
  \caption{The representative Feynman diagrams for $e^+e^- \to Zh$ in the MSSM:
(A) is the tree level diagram, and (B-D) are the self-energy, triangle and box diagrams,
respectively. }
 \label{zh}
\end{figure}

The one-loop corrections for $e^+e^- \to Zh$ in the SM and MSSM have been studied in \cite{effective-potential}.  In Fig. \ref{zh} we show the typical Feynman diagrams for the $e^+e^- \to Zh$ production in the MSSM.

The complete one-loop corrections to the process $e^+e^- \to Zh$ include two parts:
virtual corrections and real photon radiations. The virtual corrections
include a set of self-energy corrections, the vertex corrections of $eeZ$, $ZZh$ and $ZAh$,
and the box diagrams. We adopt the dimensional regularization and the constrained differential
renormalization (CDR) \cite{cdr} to regulate the ultraviolet divergence (UV) in the loop
amplitudes for the SM and MSSM, respectively. These UV singularities are removed by using
the on-shell renormalization scheme. We take the definitions of the scalar and tensor
two-, three- and four-integral functions presented in \cite{denner} and use
Passarino-Veltman method to reduce the $N$-point tensor functions to scalar
integrals \cite{Passarino-Veltman}.

Due to the infrared (IR) singularities in the vertex corrections to $e^+e^- \to Zh$,
the real photon radiation corrections should be taken into account. These IR divergences
can be canceled with the real photon bremsstrahlung corrections in the soft photon limit
by the Kinoshita-Lee-Nauenberg theorem \cite{kln}. According to the energy of the photon
$E_{\gamma}$, we split the phase space into a soft region ($E_\gamma<\Delta E_\gamma \ll \sqrt{s}/2$)
and a hard region ($E_\gamma<\Delta E_\gamma \gg \sqrt{s}/2$), where $\Delta E_\gamma$ is the
energy cut-off of the soft photon. We use the soft photon approximation formula to obtain the
soft part of the cross section \cite{soft-photon-approx} and give a fictitious mass $m_\gamma$
to the photon to eliminate the IR divergence. It should be noted that the dependence of the
real corrections on $m_\gamma$ is exactly canceled by the corresponding virtual corrections.
In the hard region, we use the well-known VEGAS \cite{vegas} routine to evaluate the cross
section. We checked that our results are independent of $m_\gamma$ and $\Delta E_\gamma$.

\subsection{Calculations for $\gamma\gamma \to h$}
The leading order $\gamma\gamma \to h$ occurs at one-loop level, where the photon beam is generated
by the backward Compton scattering of the incident electron- and the laser-beam. The number of
events is obtained by convoluting the cross section of $\gamma\gamma$ collision with the photon
beam luminosity distribution given by
\begin{eqnarray}
N_{\gamma \gamma \to h}&=&\int d\sqrt{s_{\gamma\gamma}}
  \frac{d\cal L_{\gamma\gamma}}{d\sqrt{s_{\gamma\gamma}}}
  \hat{\sigma}_{\gamma \gamma \to h}(s_{\gamma\gamma})
  \equiv{\cal L}_{e^{+}e^{-}}\sigma_{\gamma \gamma \to h}(s)
\end{eqnarray}
where $d{\cal L}_{\gamma\gamma}$/$d\sqrt{s}_{\gamma\gamma}$ is the photon-beam luminosity
distribution and $\sigma_{\gamma \gamma \to h} (s)$ ($s$ is the squared center-of-mass energy
of $e^{+}e^{-}$ collision) is defined as the effective cross section of $\gamma \gamma \to h$.
In the optimal case, it can be written as \cite{photon_collider}
\begin{eqnarray}
\sigma_{\gamma \gamma \to h}(s)&=&
  \int_{\sqrt{a}}^{x_{max}}2zdz\hat{\sigma}_{\gamma \gamma \to h}
  (s_{\gamma\gamma}=z^2s) \int_{z^{2/x_{max}}}^{x_{max}}\frac{dx}{x}
 F_{\gamma/e}(x)F_{\gamma/e}(\frac{z^{2}}{x})
\end{eqnarray}
where $F_{\gamma/e}$ denotes the energy spectrum of the back-scattered photon for the
unpolarized initial electron and laser photon beams given by
\begin{eqnarray}
F_{\gamma/e}(x)&=&\frac{1}{D(\xi)}\left[1-x+\frac{1}{1-x}-\frac{4x}{\xi(1-x)}
  +\frac{4x^{2}}{\xi^{2}(1-x)^{2}}\right]
\end{eqnarray}
with
\begin{eqnarray}
D(\xi)&=&(1-\frac{4}{\xi}-\frac{8}{\xi^{2}})\ln(1+\xi)
  +\frac{1}{2}+\frac{8}{\xi}-\frac{1}{2(1+\xi)^{2}}.
\end{eqnarray}
Here $\xi=4E_{e}E_{0}/m_{e}^{2}$ ($E_{e}$ is the incident
electron energy and $E_{0}$ is the initial laser photon energy)
and $x=E/E_{0}$ with $E$ being the energy of the scattered photon
moving along the initial electron direction.

In the calculations of $e^+e^- \to Zh$ and $\gamma\gamma \to h \to b\bar{b}$,
we use the package \textsf{FeynHiggs} to obtain the masses of the Higgs bosons
in the MSSM. By evaluating loop corrections to the $h$, $H$ and $hH$-mixing propagators,
we can determine the masses of the two $CP$-even Higgs bosons $m_h$ and $m_H$ as the
poles of this propagator matrix, which are given by the solution of
\begin{eqnarray}
  [q^2-m^{2,tree}_h+\hat{\Sigma}_{hh}(q^2)][q^2-m^{2,tree}_h+\hat{\Sigma}_{HH}(q^2)]-[\hat{\Sigma}_{hH}(q^2)]^2=0
\end{eqnarray}
where $\hat{\Sigma}_{hH}(q^2)$, $\hat{\Sigma}_{HH}(q^2)$ and $\hat{\Sigma}_{hH}(q^2)$ denote
the renormalized Higgs boson self-energies. It should be noted that
since the Higgs field renormalization constants are given in the $\overline{DR}$
scheme \cite{Complex-scheme} in \textsf{FeynArts-3.9}, we adopt the finite wave function
normalization factors $\hat{Z}_{ij}$ to ensure the correct on-shell properties of the external
particles in the $S$-matrix elements. The values of $\hat{Z}_{ij}$ can be numerically obtained
by using the package \textsf{FeynHiggs}.

Also, it should be noted that normally one has to use the tree-level Higgs masses in the whole
loop calculations to keep the gauge invariance,
while for the phase space integration we need to express the matrix element in terms of
the physical masses for the external final-states.
In our study, we take the loop-corrected Higgs boson mass as the physical mass and
adopt the proposed way in \cite{zerwas} to technically deal with this problem.
To be specific, since the tree-level process $e^+e^- \to Zh$ does not involve the
exchange of the light Higgs boson $h$, we only need to use tree-level Higgs masses
in the loop integral calculation but keep the loop-corrected mass in the phase space
integration.

\section{Numerical results and discussions}
The SM input parameters are taken as \cite{pdg}
\begin{eqnarray}
&& m_t = 172.00{\rm ~GeV}, ~m_{Z} =91.1876 {\rm ~GeV},\nonumber \\
&& \alpha(m_Z) = 1/127.9,~ \sin^{2}\theta_W = 0.231,~ \alpha_{s}(m_Z)=0.1185.
\end{eqnarray}
We define the following ratio to quantitatively show the SUSY effect in the Higgs productions
\begin{eqnarray}
  \frac{\Delta \sigma^{SUSY}}{\sigma_{SM}}=\frac{\sigma^{SUSY}-\sigma^{SM}}{\sigma^{SM}}
\end{eqnarray}
where $\sigma^{SUSY}$ and $\sigma^{SM}$ are the one-loop cross sections in the MSSM and SM, respectively.

\subsection{Results for $e^+e^- \to Zh$ in MSSM and CMSSM}
\begin{figure}[t]
 \includegraphics[width=16.5cm]{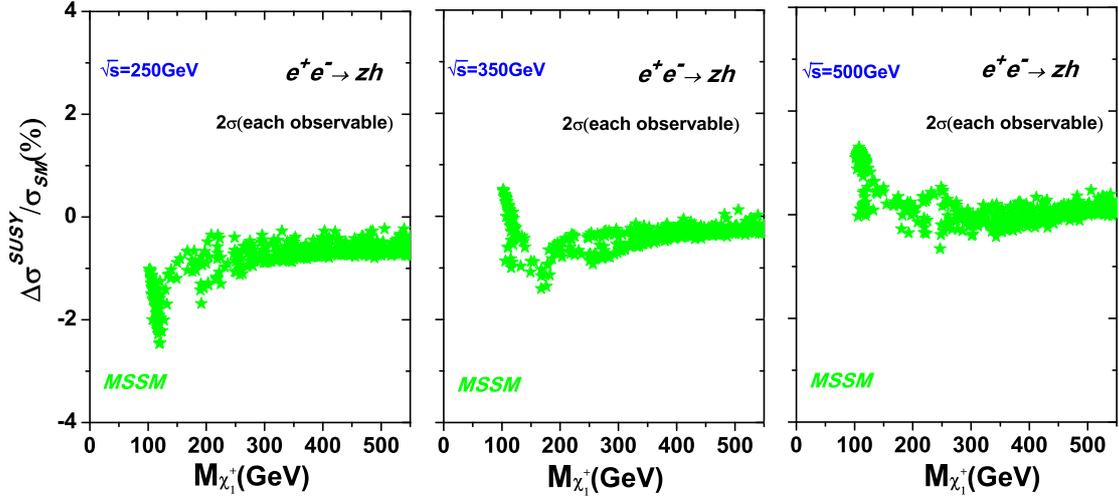}
\vspace*{-1.cm}
  \caption{The MSSM corrections to the process $e^+e^- \to Zh$ for the samples allowed by
constraints (1)-(6) at $2\sigma$ level for an $e^+e^-$ collider with $\sqrt{s}=250,350,500$ GeV.}
  \label{fig-rel-mssm-ee}
\end{figure}

In Fig.\ref{fig-rel-mssm-ee} we show
the dependence of SUSY corrections to the process $e^+e^- \to Zh$ on the chargino mass
$m_{\tilde{\chi}^+_1}$ for the samples allowed by constraints (1)-(6) at $2\sigma$ level
for an $e^+e^-$ collider with $\sqrt{s}=250,350,500$ GeV.
From this figure we can see that the SUSY corrections can be negative or positive, depending
on the masses of the sparticles in the loops and the collider energies.
For $\sqrt{s}=250$ GeV, the SUSY corrections can maximally reach $-2.5\%$ with
$m_{\tilde{\chi}^+_1}\sim \sqrt{s}/2$, which is caused by the resonant effects in the chargino loops.
Note that the bounds on the chargino mass from direct electroweakinos searches are still weak
for our samples since most of the survived points are dominated by the mixture of wino-higgsino.
Given the expected sensitivity of a 250 GeV $e^+e^-$ collider like CEPC, the residual SUSY effects
in $e^+e^- \to Zh$ can still be detected if $m_{\tilde{\chi}^{+}_{1}}<400$ GeV when the luminosity reach
about 10,000 fb$^{-1}$ \cite{higgs-report}.
Since the $hZZ$ coupling directly affects the $Zh$ production, we survey the deviation of this
coupling from the SM value and find it at most $0.05\%$. So in the MSSM the difference
of $Zh$ production from the SM is largely from the sparticle contribution.

\begin{figure}[htbp]
  \includegraphics[width=16.5cm]{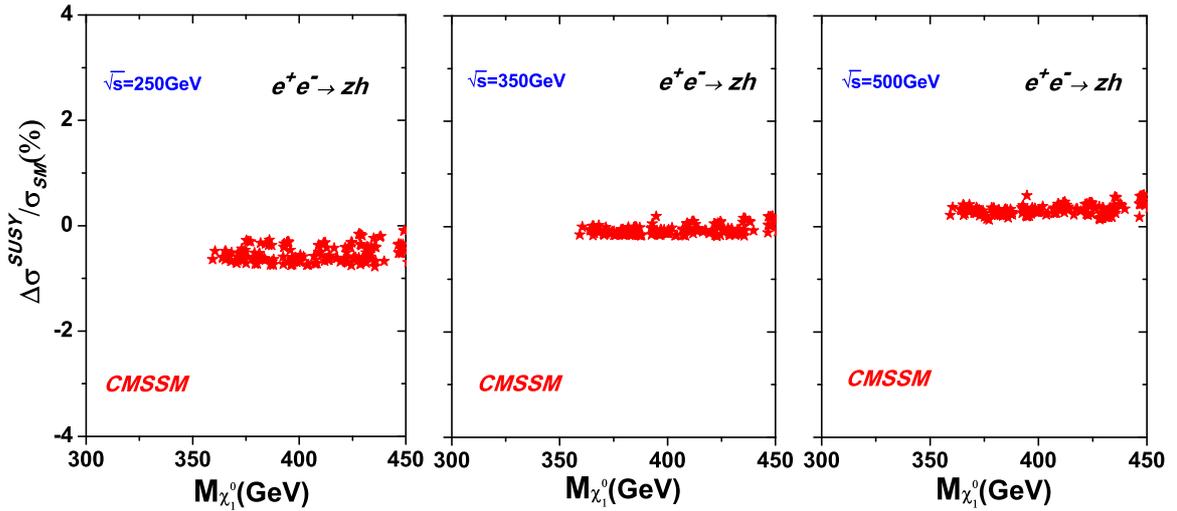}
\vspace*{-1.cm}
  \caption{Same as Fig.\ref{fig-rel-mssm-ee}, but for the CMSSM.}
  \label{fig-rel-cmssm-ee}
\end{figure}
In Fig.\ref{fig-rel-cmssm-ee} we present the dependence of SUSY corrections to the process
$e^+e^- \to Zh$ on the neutralino mass $m_{\tilde{\chi}^0_1}$ in the CMSSM with
$\sqrt{s}=250,350,500$ GeV. We find that the SUSY corrections for almost samples are
less than $0.5\%$ because the sparticles masses have been pushed up to multi-hundreds GeV region
by the inclusive sparticles searches for the CMSSM at the LHC. So it is difficult to observe
these indirect CMSSM loop effects through $e^+e^-\to Zh$ production at future $e^+e^-$ colliders.

\subsection{Results for $\gamma\gamma \to h\to b\bar{b}$ in MSSM }
\begin{figure}[htbp]
\begin{center}
\includegraphics[width=15cm]{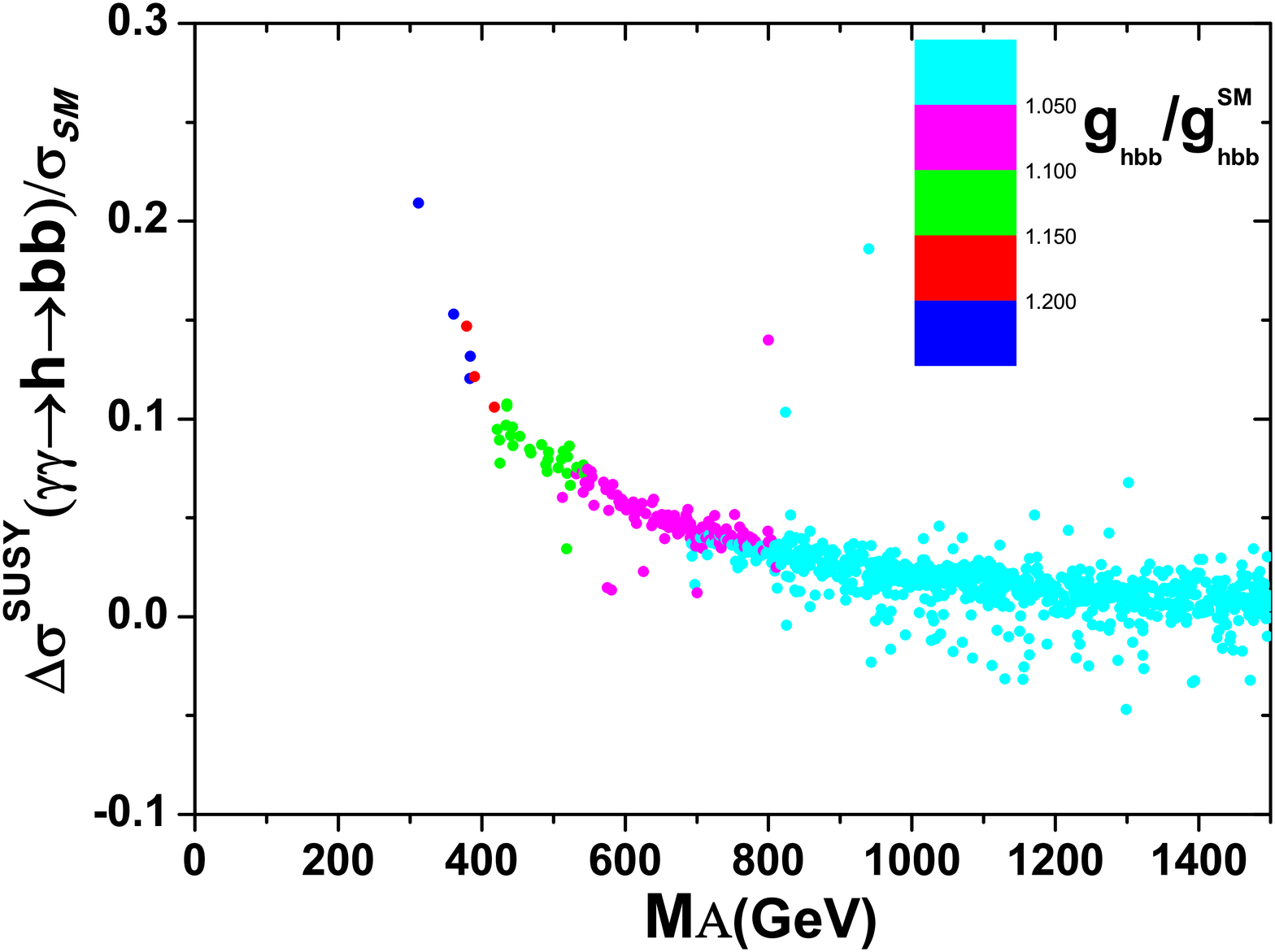}
\end{center}
\vspace*{-1cm}
\caption{The SUSY corrections to $\gamma\gamma\to h\to b\bar{b}$ at a photon collider with
center-of-mass energy above 125 GeV.}
\label{fig4}
\end{figure}

At an ILC-based photon collider, the Higgs boson can be singly produced through the photon-photon
fusion mechanism. Since the cross section of $\gamma\gamma \to h$ is proportional to the decay width
of $h \to \gamma\gamma$, the ratio
$\Delta\sigma^{SUSY}(\gamma\gamma\to h)/\sigma^{SM}(\gamma\gamma\to h)$ is independent of
the energy of ILC. Given the large branching ratio of $h \to b\bar{b}$, we calculate the
SUSY corrections to the observable $\sigma(\gamma\gamma \to h)\cdot Br(h\to b\bar{b})$
in the MSSM and display its dependence on the mass of pseudo-scalar $m_{A}$ in Fig.\ref{fig4}.

From Fig.\ref{fig4} we can see that the SUSY corrections can maximally reach $20\%$ for the
allowed samples in the small $m_A$ region with a large $\tan\beta$ due to the enhancement
of $Br(h \to b\bar{b})$. With the increase of $m_A$, the SUSY corrections drop.
Note that a light stau can make sizable loop contribution
to $\gamma\gamma \to h$, which, after considering the vacuum stability,
can enhance the cross section by a factor of 1.2.

So, if the photon-photon collison can be realized at the ILC, it will be a good place for
probing SUSY effects.
Of course, it should be mentioned that such sizable effects may be detected or further
constrained at the LHC Run-2.

\subsection{Results for $e^+e^-\to Zh_1$ and  $e^+e^-\to h A_1$  in NMSSM}
In our scan of the NMSSM parameter space we choose $h_2$ as the SM-like Higgs $h$.
In this case the lightest CP-even Higgs $h_1$ and CP-odd Higgs $A_1$ are singlet-dominant
and can be much lighter than $h$.

\begin{figure}[htbp]
\includegraphics[width=6.4in,height=3.0in]{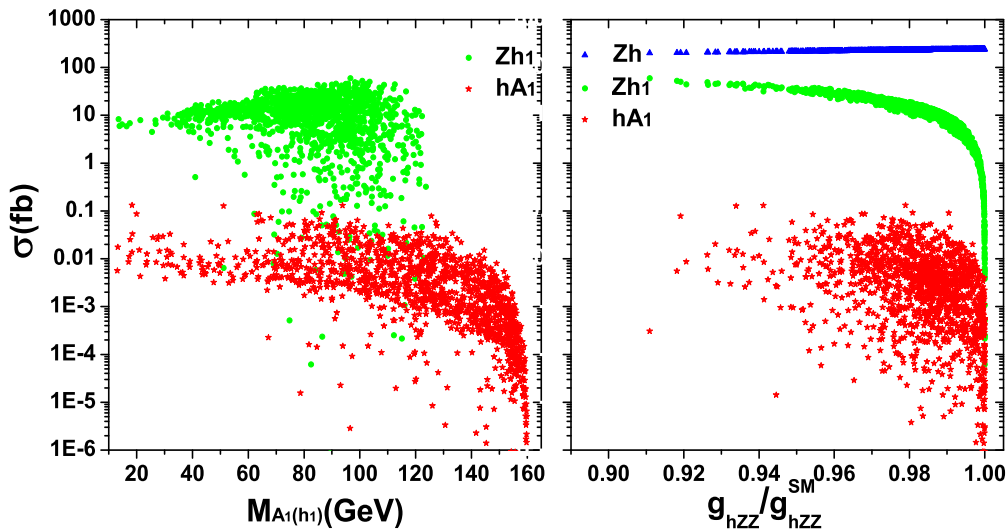}
\caption{The cross sections of $e^+e^-\rightarrow Zh_1$ and   $e^+e^-\rightarrow A_1h$
in the NMSSM for a 250 GeV  $e^+e^- $ collider.}
\label{fig5}
\end{figure}

In Fig.\ref{fig5} we show the leading-order cross sections for $e^+e^-\rightarrow Zh_1$ and
$e^+e^-\rightarrow A_1h$ in the NMSSM for $\sqrt{s}=250$ GeV.
From the right panel we see that the $e^+e^-\rightarrow Zh_1$ production rate
varies between a rather large
range and most samples give a cross section larger than 1 fb.
It is interesting that the largest production rate increases
with the mass of $h_1$. The reason is that the $e^+e^-\rightarrow Zh_1$ production rate
largely depends on the $h_1ZZ$ coupling, which comes from the mixing between the singlet and
the doublet Higgs fields. When the two masses of $h_1$ and $h$ get closer, the mixing
generally become larger and thus $e^+e^-\rightarrow Zh_1$ increases with the $h_1$ mass.
This situation is different from the $e^+e^-\rightarrow A_1h$ production, whose cross section
decreases with the increase of $A_1$ mass since the production rate would be kinematic enhanced
when $A_1$ is light.

Note that the $e^+e^-\rightarrow A_1h$ production rate is much smaller than $e^+e^-\rightarrow Zh_1$.
The largest cross section can only reach 0.1 fb. The reason is that this production rate depends on
the $ZA_1h$ coupling. This couplings arises from the mixing between the two CP-odd scalars and a
moderately large $M_{A_1}$ would induce a small mixing.

The right panel of  Fig.\ref{fig5} shows the leading-order cross sections
of three production channels versus the $hZZ$ coupling normalized to the SM value.
We see that the cross section of $e^+e^-\rightarrow Zh_1$
is sensitive to the deviation of $hZZ$ coupling from the SM value.
When the $hZZ$ coupling approaches to the SM value, the  cross section of
$e^+e^-\rightarrow Zh_1$ drops sharply.
For the channel $e^+e^-\rightarrow A_1h$,
although its production rate is much smaller than $e^+e^-\rightarrow Zh_1$,
it is not so sensitive to the $hVV$ coupling.
We numerically checked that as the $hZZ$ coupling approaches to the SM value,
the production rate of $e^+e^-\rightarrow A_1h$ can still reach 0.1 fb.

Since the dominant decay mode of the light Higgs bosons is $b\bar{b}$,
the exotic Higgs productions $e^+e^-\rightarrow A_1h$ and  $e^+e^-\rightarrow Zh_1$
will lead to $4b$ and $Z+2b$, respectively. These signals can be efficiently detected
at an $e^+e^-$ collider.  Also note that at such an $e^+e^-$ collider the loop-induced
Higgs production $e^+e^-\to h\gamma$ can get enhanced in SUSY \cite{ee-hgamma}.
All these processes can jointly serve as a good probe for SUSY models.

\section{Conclusion}
In this work we examined the SUSY residual effects in the process $e^+e^- \to Zh$
at an $e^+e^-$ collider with center-of-mass energy above 250 GeV
and $\gamma\gamma \to h \to b\bar{b}$ at a photon collider with center-of-mass energy above
125 GeV. We found that the SUSY corrections to $e^+e^- \to Zh$ can reach a few percent
in the parameter space allowed by current experiments.
The production rate of $\gamma\gamma \to h \to b\bar{b}$ can be enhanced by
a factor of 1.2 over the SM prediction. We also calculated the exotic
Higgs productions $e^+e^-\rightarrow Zh_1$ and  $e^+e^-\rightarrow A_1h$ in the NMSSM.
We found that for an $e^+e^-$ collider with center-of-mass energy of 250 GeV
the $e^+e^-\rightarrow Zh_1$ and $e^+e^-\rightarrow A_1h$ production rates
can reach 60 fb and 0.1 fb, respectively.
These processes will jointly serve as a probe of SUSY in the proposed
$e^+e^-$ collider like CEPC, TLEP or ILC.

\section*{Acknowledgement}
This work was supported  by the National Natural Science Foundation of China (NNSFC) under
grant Nos. 10821504, 11222548, 11305049 and 11135003, by Program for New
Century Excellent Talents in University, and also by the ARC Center of Excellence
for Particle Physics at the Tera-scale.

\end{document}